\documentclass[preprint,aps,prb]{revtex4-2}
\usepackage{amsmath,mathtools}
\usepackage{amssymb}
\usepackage{gensymb}
\usepackage{amsthm}
\usepackage{graphicx}
\usepackage{physics}
\usepackage {xcolor}
\newcommand{\e}{\varepsilon}
\newcommand{\bea}{\begin{eqnarray}}
\newcommand{\eea}{\end{eqnarray}}
\newcommand{\beq}{\begin{equation}}
\newcommand{\eeq}{\end{equation}}

\begin{document}
	
\title{Valley and Spin Polarized States in Bernal Bilayer Graphene}

\author{R. David Mayrhofer}
\affiliation{School of Physics and Astronomy and William I. Fine Theoretical Physics Institute, University of
Minnesota, Minneapolis, MN 55455, USA}

\author{Andrey V. Chubukov}
\affiliation{School of Physics and Astronomy and William I. Fine Theoretical Physics Institute, University of
Minnesota, Minneapolis, MN 55455, USA}

\date{\today}

\begin{abstract}
We present the results for the evolution of the Fermi surfaces under variation of number density and displacement field for spin and valley-polarized states in Bernal bilayer graphene (BBG) using a realistic form of the electronic dispersion with trigonal warping terms. Earlier studies without trigonal warping have found discrete half-metal and quarter-metal states with full spin and/or valley polarization and complete  depletion of some of the Fermi surfaces. We show that when trigonal warping terms are included in the dispersion, partially polarized states with large but non-complete polarization and with both majority and minority carriers present, emerge at small doping, as seen in the experimental data. We show the results when the intervalley and intravalley interactions are equal as well as when they are unequal.
\end{abstract}

\maketitle

\section{Introduction}
Graphene multilayers in transverse displacement fields have been shown to host a variety of exotic phases of matter. Recent experiments on Bernal bilayer graphene (BBG) and rhombohedral trilayer graphene (RTG) as well as other moir\`e-less graphene multilayers have revealed a cascade of electronic phase transitions, including valley and spin-polarized states, as well as potentially unconventional superconducting phases~\cite{zhou2021rtg,zhou2021sc,zhou2022bbg,seiler2022,*Seiler2024,barrera2022,han2023,trevol2024,han2024,holleis2025}, the latter grows in the presence of an Ising spin-orbit coupling imposed by putting  graphene multilayers in proximity with a layer of tungsten diselenide (WSe$_2$) ~\cite{zhang2023,li2024}. A rich phase diagram comes from the ability to independently tune the doping level of the system and modify the band structure via a transverse displacement field. These results are reminiscent of findings on twisted bilayer graphene, where spin/valley orders and superconductivity have also been found~\cite{cao2018,cao2018insulator,yankowitz2019,lu2019,andrei2020,cao2021}.

\begin{figure}[h]
	\begin{center}
		\includegraphics[scale=.6]{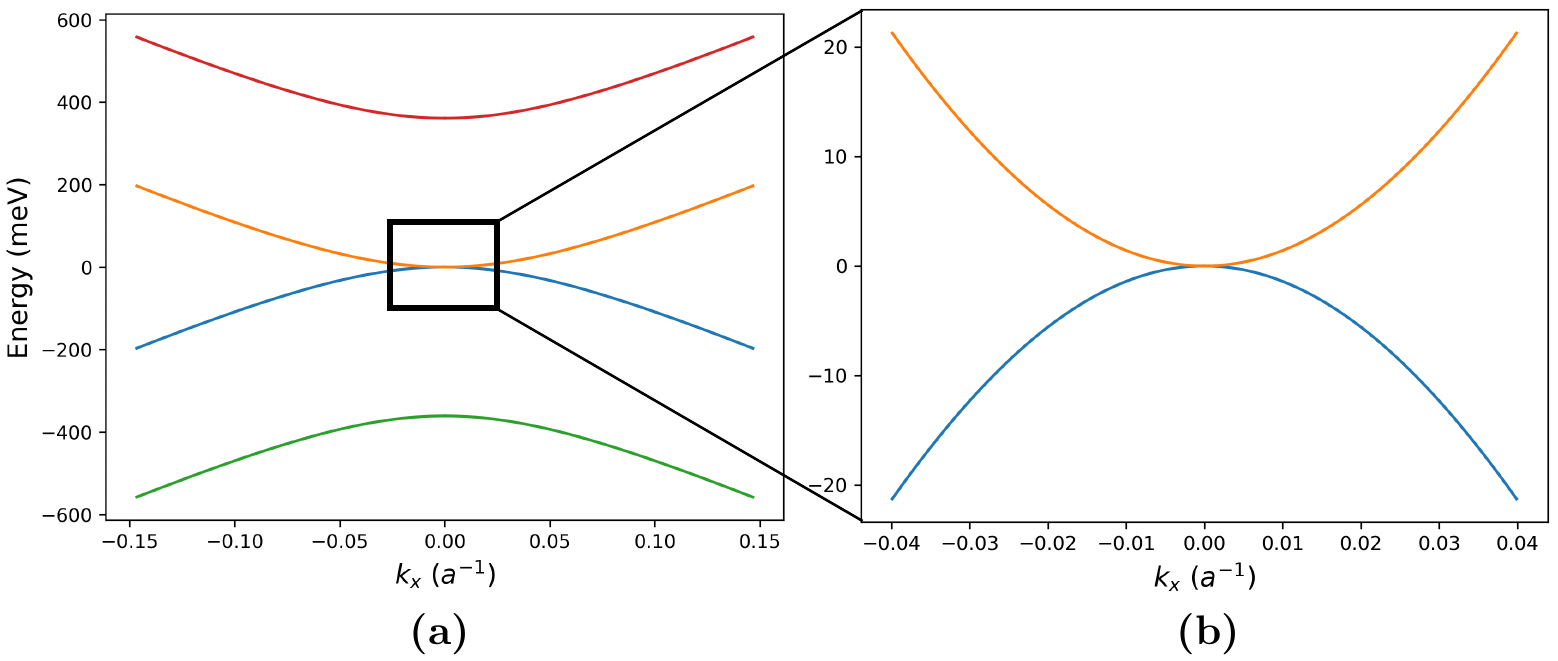}
		\caption{
  The band structure of BBG close to the K point along $k_y=0$
   in the absence of trigonal warping and displacement field.
   In (a), we show all four bands, demonstrating that the uppermost and lowermost
    ones are well separated from the middle two bands. In (b), we show
     only
      the middle
     bands. }
		\label{band_structure}
	\end{center}
\end{figure}

Some theoretical studies of BBG and RTG in a displacement field
 ~\cite{ghazaryan2021,ghazaryan2023,pantaleon2023,dong2023transformer,dong2023signatures,dong2023} focused primarily on the mechanism of superconductivity. Others~\cite{chichinadze2022,*chichinadze2022letters,haoyu2023,xie2023,lee2024,koh2024rtg,koh2024bbg,wang2024electrical,friedlan2025} aimed at understanding of spin and valley ordered normal states that border superconductivity. The electronic structure of BBG is well described by the tight-binding model and consists of four bands~\cite{mccann2006,McCann2013} (see Fig. \ref{band_structure}a). The uppermost and lowermost bands are irrelevant at low doping, as they are far away from the chemical potential. The low-energy physics  involves the two middle bands.  In the absence of the displacement field and trigonal warping (the case in Fig. \ref{band_structure}), these bands are quadratic and touch each other at special points at the Brillouin zone boundary, labeled $K$ and $K'$ (Fig. \ref{band_structure}b). The leads to a circular Fermi surface, whose size gradually increases with either hole or electron doping.

\begin{figure}[h]
	\begin{center}
		\includegraphics[scale=.6]{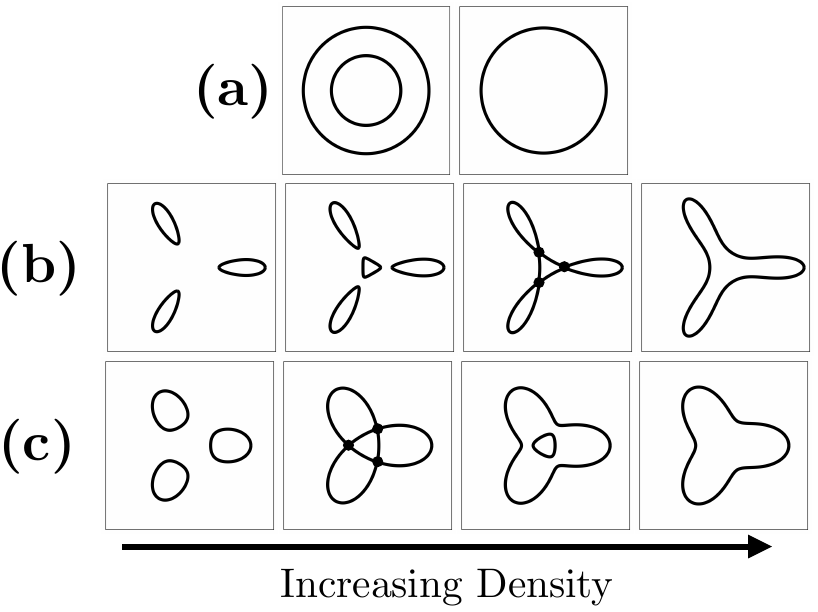}
		\caption{Schematic depictions of the evolution of the Fermi surface in BBG as particle density is increased when there is (a) no trigonal warping, but finite displacement field, (b) finite trigonal warping and low displacement field, and (c) finite trigonal warping with large displacement field. The points where the Fermi surfaces join together are marked with black dots.}
		\label{fermi_surfaces}
	\end{center}
\end{figure}

However, when one includes both a finite displacement field $D$  and trigonal warping, the Fermi surface displays more complicated behavior and its shape and topology evolve with doping~\cite{McCann2013, mccann2006asymmetry}.
Namely, at a finite $D$ and still zero trigonal warping, the Fermi surface emerges with a finite $k_F$ at infinitesimal doping and becomes annular at a finite doping, with the inner Fermi surface shrinking as doping increases and eventually disappearing. (Fig. \ref{fermi_surfaces}a).
When both $D$ and trigonal warping terms in the dispersion are non-zero, the Fermi surface at the lowest doping (electron doping  for definiteness) consists of three small pockets near $K$ and three near $K'$ (Fig. \ref{fermi_surfaces}b).
As doping is increased, the Fermi surfaces evolve, and the type of the evolution depends on the ratio of the displacement field and the hopping responsible for trigonal warping. When the ratio is small, a fourth pocket forms at some small doping, and at a larger doping touches the other three pockets, leading to a Van Hove singularity in the density of states (Fig. \ref{fermi_surfaces}b).
At even larger doping, a single larger pocket is formed. When the ratio exceeds a certain threshold, no new pocket is formed at small doping. Instead, three electron pockets gradually approach each other as doping increases, and touch each other at some doping, again leading to a Van Hove singularity (Fig. \ref{fermi_surfaces}c). At larger doping, the Fermi surface consists of a larger electron pocket and a smaller hole pocket, which eventually closes as doping continues to increase.

We show that the combination of  trigonal warping {\it and}  a finite displacement field enhances the density of states by  a large factor -- the ratio of the intra-layer hopping and the one responsible for trigonal warping.
This enhancement is not confined to the Van Hove singularity and is present in a wide range of dopings (see Fig. \ref{dos}), where, as we show below, the system develops spin/valley orders. This enhancement  develops already at a small displacement field and becomes parametrically large  for the range of displacement fields where the van Hove singularity is towards an annular Fermi surface (see below). A large density of states  enhances the effects of the electron-electron interactions and should generally promote the formation of ordered states.

To study these states, a common approach is to consider a two valley, two spin model, where electrons located in the vicinity of either the $K$ or $K'$ points and interact via a screened electron-electron repulsive interaction.
We follow Refs.~\cite{dong2023,chichinadze2022,raines2024isospin,*raines2024unconventional,raines2024stoner} and approximate the relevant interactions by three constants, $U_1, U_2$, and $U_3$, where $U_1$ and $U_2$ describe the interactions between fermionic densities in the same valley and different valleys, respectively, and $U_3$ is the exchange interaction (intervalley scattering). Within this framework, there is the potential for five different types of order parameters to develop. These are valley polarization (VP), spin polarization in the K or K' valley (SP1/SP2), charge inter-valley coherence (IVC), and spin inter-valley coherence (sIVC).
The first three are $q=0$ orders, the other two are orders with momenta $Q = K-K'$.
In the specific scenario where the intravalley and intervalley interactions terms are equal and there is no intervalley scattering, all of these orders occur at the same point, and the order parameters furnish the adjoint representation of the $SU(4)$ group \cite{chichinadze2022,*chichinadze2022letters}. However, one can set the intravalley and intervalley interactions terms to be inequal and the intervalley scattering to be finite, in which case the $SU(4)$ symmetry is broken and the orders occur at different interaction strengths~\cite{dong2023}.
We note in this regard that recent works~\cite{calvera2024nematicity,raines2024stoner,Huang2025} have found that, even if one starts with a fully $SU(4)$ symmetric model at a mean-field level, going beyond the mean field approximation produces effective interactions that break the $SU(4)$ symmetry.

     \begin{figure}[h]
	\begin{center}
		\includegraphics[scale=.6]{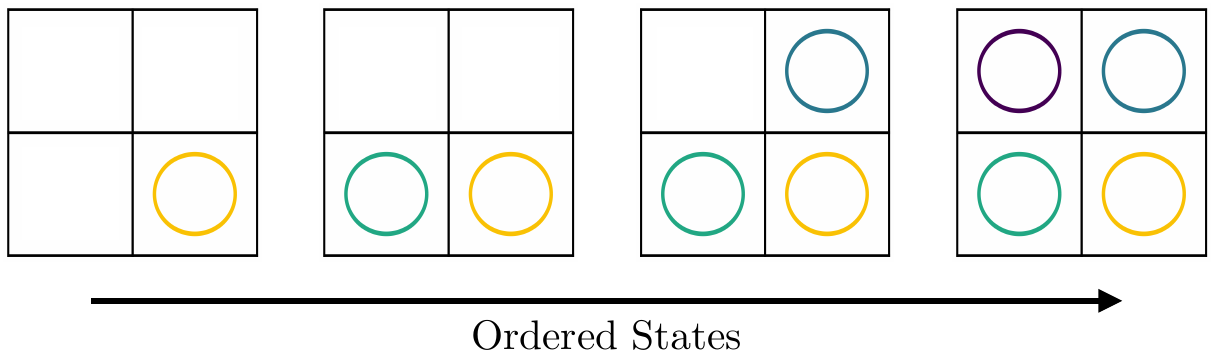}
		\caption{General evolution of the Fermi surfaces in each isospin in an isotropic system with two spins and two valleys, as studied in Ref. \cite{raines2024isospin,*raines2024unconventional}. }
		\label{isotropic_fermi_surfaces}
	\end{center}
\end{figure}
     The $q=0$ ordered states in a two valley, two spin system with an isotropic dispersion and circular Fermi surfaces (no trigonal warping)  have been studied in \cite{chichinadze2022,*chichinadze2022letters,dong2023,raines2024isospin,*raines2024unconventional}.
      These authors modeled the softening of the dispersion in the presence of a displacement field by replacing $k^2$ by a weaker  dispersion $k^{2\alpha}$ with $\alpha$ between 1 and 2. They found a set of the ordered states, which all are fully polarized in spin and/or valley subspaces. The feedback from such orders on fermions yields half-metal and quarter-metal states (and, in one limit, also three-quarter-metal state) in which one or more Fermi surfaces get fully depleted.
We depict these states in Fig. \ref{isotropic_fermi_surfaces}.
Within this approximation, the ordering develops at arbitrary small doping and exists over some doping range.
The electronic structure changes with increasing doping from a quarter-metal to a half-metal, to a three-quarter metal and finally to a full metal, when order disappears.
 Finite $q$  IVC and sIVC orders have been primarily studied in twisted bilayer graphene (TBG) ~\cite{Chichinadze2020,Kwan2021,Kwan2024,Wang2023,Wang2024} with the key focus on Kekule-type order, and in un-twisted rhombohedral trilayer graphene (RTG)~\cite{Chatterjee2022,Vituri2024}
  In both systems, an IVC order has been determined experimentally (see Refs.~\cite{Nuckolls2023,Kim2023} for TBG and Refs. \cite{Arp2024,Liu2024,Auerbach2025} for RTG).

\begin{figure}[h]
	\begin{center}
		\includegraphics[scale=.6]{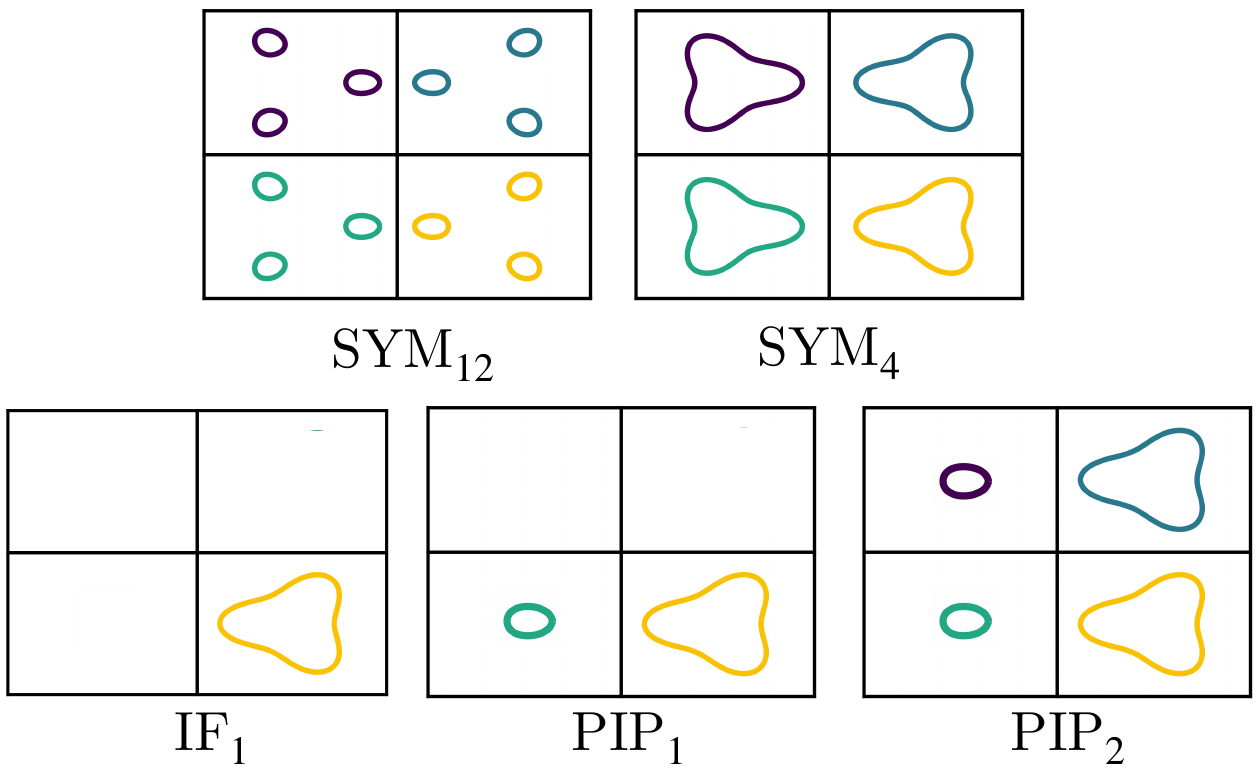}
		\caption{Depiction of the Fermi surfaces found in experiment, where the labels are taken from  Ref. \cite{zhou2022bbg}. These Fermi surfaces were originally determined from the quantum oscillation data, assuming that both valley and spin remain good quantum numbers}
		\label{pip}
	\end{center}
\end{figure}

In our analysis, we focus primary on the understanding of spin/valley ordered states in BBG. The electronic structure of BBG at a given displacement field $D$ and doping $x$  has been extracted from quantum oscillations and from the measurements of the electronic compressibility.   The resulting phase diagram in the $(D,x)$ plane has been presented in~~\cite{zhou2022bbg,zhang2023,seiler2022,*Seiler2024,holleis2025}. It is similar, although not identical, to the one reported for rhombohedral trilayer graphene~\cite{zhou2021rtg,trevol2024,Arp2024,Liu2024,Auerbach2025}. In particular, to the best of our knowledge, no evidence for incommensurate IVC states has been reported.
A number of ordered states have been detected from the data as evidenced  from the divergences of the compressibility along several lines in the $(D,x)$ plane and the changes in the number and the shapes of the Fermi surfaces, extracted from Shubnikov-de Haas oscillations of resistivity.   To zeroth approximation, the ordered states can be characterized as half-metal and quarter-metal states
(a half-metal state is likely spin polarized~\cite{barrera2022,seiler2022}).
However, there are two nuances: (i) ordering develops at a finite doping, and (ii)  in addition to states with two or one larger Fermi surfaces (half-metal and quarter-metal states), which naturally emerge from fully polarized spin/valley orders, there are also states in which both larger and smaller Fermi surfaces are present.  We depict the states, reported in  the experiments in Fig. \ref{pip}.
We will be using the notations from Ref. \cite{zhou2022bbg}. More detailed notations have been introduced in Ref. \cite{patterson2024}. The quarter-metal state with the full polarization has been labeled there as IF$_1$ and the states with both larger and smaller Fermi surfaces as PIP$_1$ and PIP$_2$. The last  two states are close to  quarter-metal and half-metal, respectively, but it is natural to expect that in these states spin/valley polarization is only partial.
The question then becomes if a system  described by the full tight-binding model with hopping responsible for the trigonal warping can produce partially polarized states, and, if so, how strongly does the trigonal warping terms affect the presence of these partially polarized states.

In this communication we address this issue. We  consider  $q=0$ orders with valley and/or spin polarization (VP and SP1/SP2). We follow earlier works~~\cite{calvera2024nematicity,raines2024stoner,raines2024isospin,*raines2024unconventional,Huang2025} and  neglect the exchange interaction $U_3$ as it is smaller than $U_1$ and $U_2$ and does not qualitatively affects  $q=0$ orders.  We apply the ladder  approximation, as in earlier works, to detect and classify the orders,  but use the  full dispersion of fermions, including trigonal warping terms. We explore the effects of trigonal warping and $SU(4)$ symmetry breaking on the occupation of each isospin and present the phase diagrams for the system as number density and displacement field are varied.

We find that trigonal warping strongly effects the structure of the polarized states.
More specifically,  there still exist fully polarized half-metal and quarter-metal states, but in some doping ranges new  states appear, in which  the majority carriers  are accompanied by minority carriers.
For $U_1 =U_2$, we find doping ranges of IF$_1$ and PIP$_2$ states, while for $U_1 \neq U_2$, we found the ranges for all three states IF$_1$, PIP$_1$ and PIP$_2$, seen in the experiments~~\cite{zhou2022bbg,zhang2023,seiler2022,*Seiler2024,holleis2025}.
Additionally, for both equal and non-equal values of  $U_1$ and $U_2$, we find a non-ordered state at the smallest doings. This again agrees with the experiments.

The rest of the paper is organized as follows. In Sec. \ref{background}, we present the single-particle Hamiltonian in BBG, obtain the band structure, and derive self-consistent equations for spin and valley order parameters.
We also obtain the density of states and demonstrate its enhancement under the presence of a displacement field.
In Sec. \ref{results}, we first present the phase diagram of the system neglecting trigonal warping, both when the interaction is $SU(4)$ symmetric and when $SU(4)$ symmetry is broken.  We then reintroduce the trigonal warping to the system, obtain the phase diagrams again  for the case when spin and valley instabilities are degenerate and when the symmetry is broken, and compare them with the case of no trigonal warping~\footnote{There is no exact $SU(4)$ symmetry in this case as, due to trigonal warping, the polarization bubbles at zero momentum and at momentum $Q=K-K'$ are not equivalent.}.
We compare the isospin occupations in each case with quantum oscillation data.
We present our conclusions in Sec. \ref{conclusions}.

\section{Background and Formalism}
\label{background}
\subsection{BBG Model Hamiltonian}
\begin{figure}[h]
	\begin{center}
		\includegraphics[scale=.9]{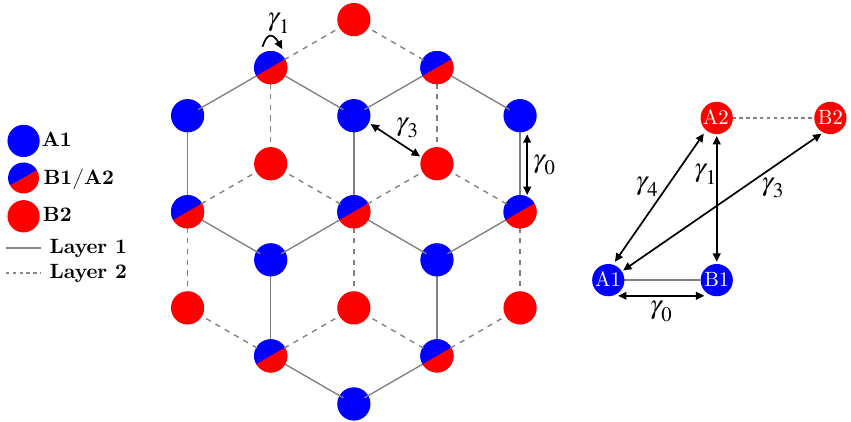}
		\caption{Illustration of the BBG lattice from the viewed from the top (left) and from the side (right). The tunneling parameters have been labeled on the appropriate bonds.}
		\label{tunneling_parameters}
	\end{center}
\end{figure}

To model the single-particle Hamiltonian of BBG, a common approach is to consider the tight-binding model on two stacked hexagonal lattice (Fig. \ref{tunneling_parameters}).
This model matches well with the \textit{ab initio} calculations \cite{jeil2014}, so we will adopt it. The single particle Hamiltonian for the system can be expressed as follows,
\begin{align}
\label{single_particle} H_{s} = \sum_{\vb k,\alpha} \begin{pmatrix} a^{\dagger}_{1,\vb k,\alpha} &b^{\dagger}_{1,\vb k,\alpha} &a^{\dagger}_{2,\vb k,\alpha} &b^{\dagger}_{2,\vb k,\alpha} \end{pmatrix}
\begin{pmatrix}
D/2 & \gamma_0 f(\vb k)  & -\gamma_4 f(\vb k) & -\gamma_3 f^*(\vb k) \\
\gamma_0 f^*(\vb k) &  \delta + D/2 & \gamma_1 & -\gamma_4  f(\vb k) \\
-\gamma_4 f^*(\vb k)  & \gamma_1 & \delta -D/2 & \gamma_0 f(\vb k)   \\
-\gamma_3 f(\vb k)& -\gamma_4 f^*(\vb k)& \gamma_0 f^*(\vb k) & -D/2
\end{pmatrix}
\begin{pmatrix}
a_{1,\vb k,\alpha}\\
b_{1,\vb k,\alpha} \\
a_{2,\vb k,\alpha} \\
b_{2,\vb k,\alpha}
\end{pmatrix},
\end{align}
where $a_{1(2)}$ and $b_{1(2)}$ respectively are the annihilation operators on the A and B sublattices of layer 1(2), $D$ is the displacement field, $\delta$ is the AB-sublattice potential difference, and $f(\vb k)$ is the form-factor for the nearest-neighbor hopping within a hexagonal lattice. The $\gamma_i$ are the hopping parameters, with $\gamma_0$ describing the intralayer hopping between the $A1(2)$ and $B1(2)$ sites, $\gamma_1$ describing hopping between $A2$ and $B1$ sites, i.e. the sites directly on top of one another, $\gamma_3$ describing hopping between the $A1$ and $B2$ sites, and $\gamma_4$ between $A(B)1$ and $A(B)2$. We depict these tunneling parameters graphically in Fig. \ref{tunneling_parameters}.
For practical calculations, we will be using the values for the $\gamma_i$ and $\delta$ obtained from \textit{ab initio} calculations in Ref. \cite{jeil2014}:
\begin{align}
\label{parameters} \gamma_0 = 2610 \text{meV}, \, \gamma_1 = 361 \text{meV}, \, \gamma_3 = 283 \text{meV}, \, \gamma_4 = 138 \text{meV}, \delta = 15 \text{meV}.
\end{align}

Close to the $K,K'$ points, we can write $f(\vb k)$ as
\begin{align}
f(\vb k) = \frac{\sqrt{3} a}{2}\left(\tau k_x - i k_y \right)
\end{align}
where $a = 2.46$\AA$ $ is the lattice constant, $\tau = \pm 1$ is the valley index, and $\vb k$ is centered at either the $K$ or $K'$.

\begin{figure}[h]
	\begin{center}
		\includegraphics[scale=.7]{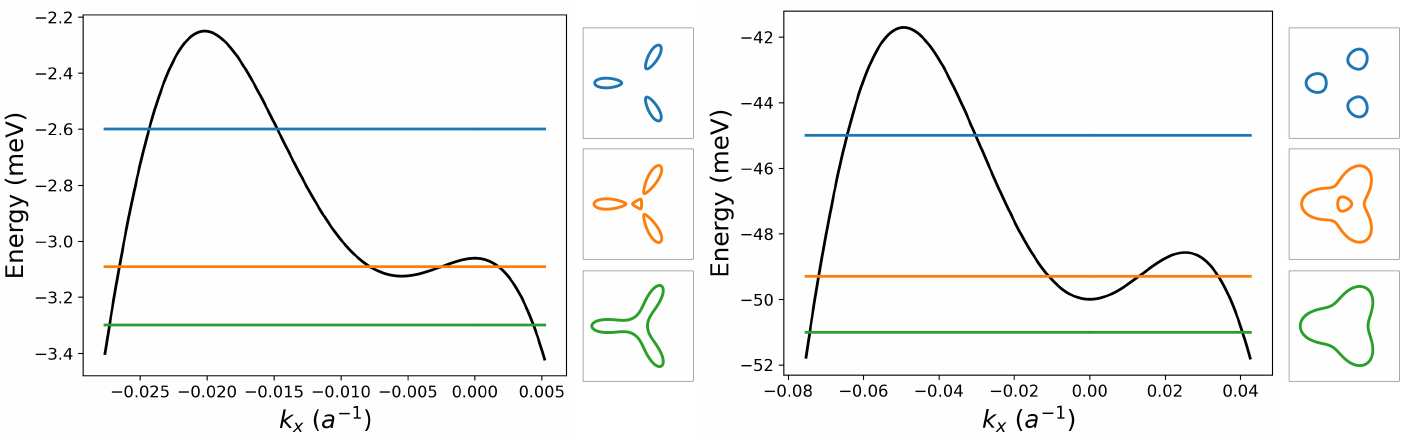}
		\caption{The hole band plotted along the $k_x$ axis for $D<D_c$ (left) and $D>D_c$ (right), along with Fermi surfaces for several different chemical potentials.
The difference between $D < D_c$ and $D>D_c$ is in that in the first case the dispersion has a local maxium at $k_x =0$, while in the second it has a local minimum.
For completeness, we kept $\gamma_4$ and $\delta$ in the dispersion (the two terms that account
    for particle-hole asymmetry), but the effects of these terms are truly small~
      \protect\cite{McCann2013}.}
		\label{one_band}
	\end{center}
\end{figure}

The band dispersion is obtained by diagonalizing $4 \times 4$ matrix in Eq. \ref{single_particle}.
This dispersion can be written in a tractable form when $\gamma_4 = \delta = 0$~\cite{McCann2013}. In this limit, it is particle-hole symmetric and the energies of the four bands are $\pm |\e_{\vb k,\tau}|$, where
\begin{align}
\label{analytic_solution} \e_{\vb k,\tau}^2 =& \frac{\gamma_1^2}{2} + \frac{D^2}{4} + \left(\gamma_0^2 + \frac{\gamma_3^2}{2} \right) |f(\vb k )|^2 \pm \sqrt{\Gamma},\\
\nonumber \Gamma =& \frac{1}{4} \left( \gamma_1^2 - \gamma_3^2 |f(\vb k)|^2\right)^2 + \gamma_0^2 |f(\vb k)|^2 \left( \gamma_1^2 + D^2 + \gamma_3^2 |f(\vb k)|^2 \right) \\
 \nonumber & - \gamma_0^2 \gamma_1 \gamma_3 \left( f(\vb k )^3 +f^*(\vb k )^3 \right).
\end{align}
The two outer bands are obtained by taking $+\sqrt{\Gamma}$, and the two middle bands by taking $-\sqrt{\Gamma}$.
Expanding $\Gamma$ in $k$, using that $\gamma_3  \ll \gamma_0$, we find
\begin{align}
\label{dispersion} \e_{\vb k,\tau}^2
\approx
\left( \frac{ \gamma_1 \gamma_3}{\gamma_0} \right)^2 \left( \frac{\beta^2}{4} + \frac{\gamma_3^2}{\gamma_0^2}  z^2 \left(1 - \beta^2 + 2z \tau \cos 3 \theta \right) \right),
\end{align}
where $\theta$ is the angle between $\vb k$ and the $x-$axis, and we introduced $\beta = \frac{D \gamma_0}{\gamma_1 \gamma_3}$, $z = \frac{\sqrt{3}}{2} \frac{\gamma_0^2}{\gamma_1 \gamma_3} ak $.
The value of $\beta$ determines the different scenarios for the Fermi surface evolution with doping, depicted in Figs. \ref{fermi_surfaces}b and \ref{fermi_surfaces}c. For $\beta <1$, the fourth pocket gets formed at small doping, and the Fermi surface evolution is as in  Fig. \ref{fermi_surfaces}b. When $\beta >1$, the Fermi surface evolution is as in Fig. \ref{fermi_surfaces}c.  The critical displacement field, separating the two cases, is $D_c = \frac{\gamma_1 \gamma_3}{\gamma_0}$. We plot the dispersion for $\tau = +1$ along the $k_x$ direction in Fig. \ref{one_band} for the hole band, the one relevant for experiments.  

In Fig. \ref{one_band}, the band is asymmetric with respect to positive and negative $k_x$. This comes from the angular dependence of the dispersion as seen in Eq. (\ref{dispersion}). The behavior of the bands is dependent on $D$. If $D<D_c$, then the three Fermi pockets will join together with the central pocket along $\theta = \pi/3, \pi,$ and $5\pi/3$. In the band structure, this corresponds to the chemical potential crossing the local minimum in the band. If $D>D_c$ however, the three pockets will touch along $\theta = 0, 2\pi/3$, and $4 \pi/3$. This corresponds to the chemical potential crossing the local maximum in the band structure that occurs along these directions. If $\tau=-1$, the situation is the same, but with the angle rotated by $\pi/3$.

\begin{figure}[h]
	\begin{center}
		\includegraphics[scale=.6]{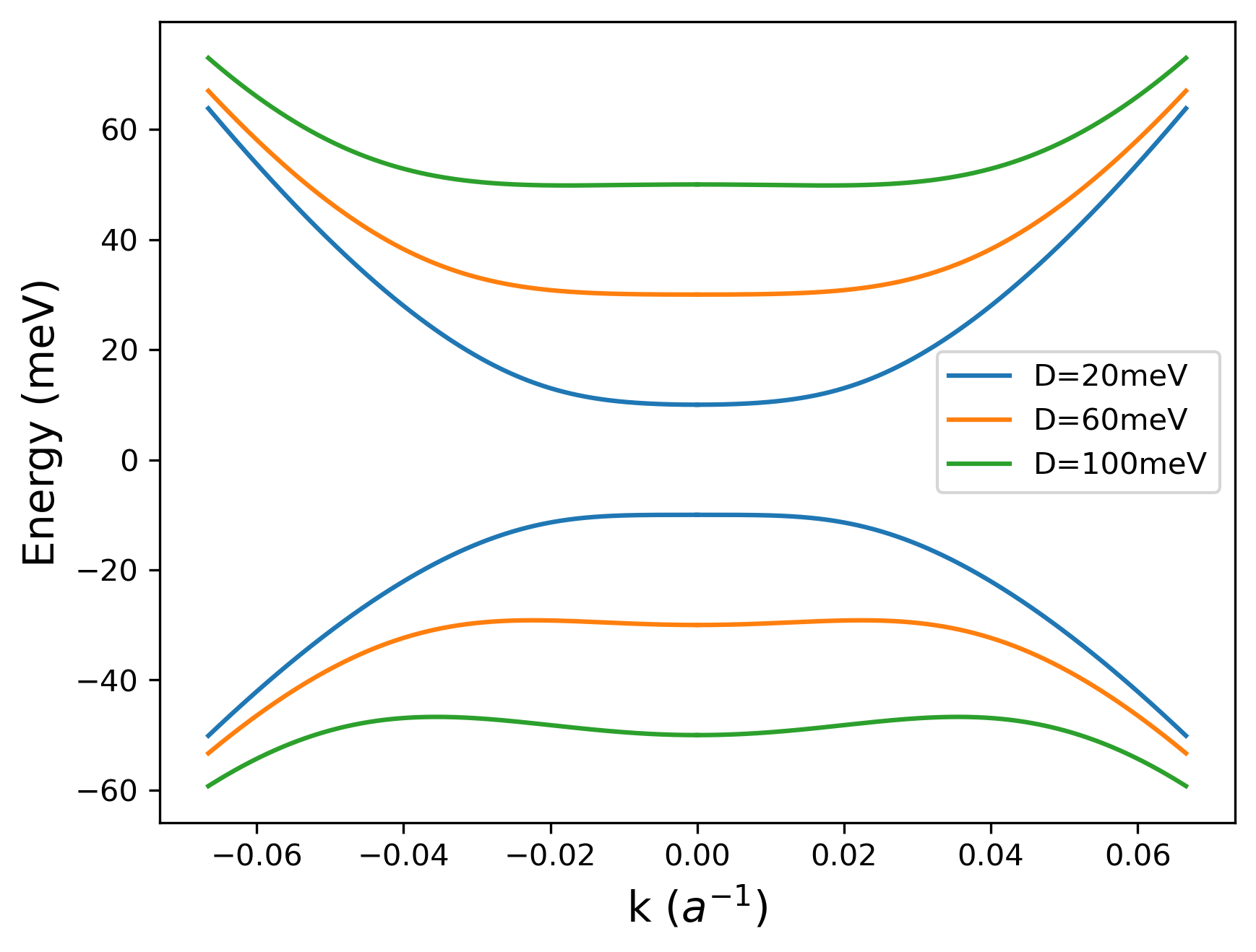}
		\caption{The particle and hole bands plotted along $\theta = \pi/6$ as defined in the text.}
		\label{flat_bands}
	\end{center}
\end{figure}

We also note that, when $\theta = \pi/6$, the cubic term will also vanish. One can then see that, when $\beta = 1$ and $\theta = \pi/6$, the leading order momentum dependent term will be quartic, and the band is maximally flat. We plot the bands along this line $\theta = \pi/6$ in Fig. \ref{flat_bands}.

When determining the ordered states of the system, we take finite $\gamma_4$ and $\delta$. The band that is relevant for recent experiments is the larger of the two hole bands, so we restrict our analysis to include only contributions from this energy. Therefore, we can write out the effective single Hamiltonian for the system as follows,
\begin{align}
H_{s} = \sum_{\vb k,\tau,\alpha} \e_{\vb k,\tau}c^{\dagger}_{\vb k,\tau,\alpha} c_{\vb k,\tau,\alpha},
\end{align}
where $\e_{\vb k,\tau}$ is the dispersion near the $K,K'$ point of the hole band, and $c_{\vb k,\tau,\alpha}$ is the annihilation operator in the diagonalized basis. In the limit of $\gamma_4=\delta=0$, the above $\e_{\vb{k},\tau}$ will reduce to the expression found in Eq. (\ref{analytic_solution}).

\subsection{Density of States}

We can estimate the value of the density of states by using the analytic expression for the dispersion in Eq. (\ref{dispersion}). We have
\begin{align}
N_F &= 4\int \frac{d^2k}{(2\pi)^2} \delta(\mu - \e_k)\\
\nonumber &= 4\left(\frac{\gamma_1 \gamma_3}{\gamma_0^2 b}\right)^2\int
\frac{z dz d \theta}{(2\pi)^2}
 \delta\left(\mu - \frac{ \gamma_1 \gamma_3}{\gamma_0} \left[ \frac{\beta^2}{4} +\frac{\gamma_3^2}{\gamma_0^2}z^2 \left( 1 - \beta^2 - 2 z \cos 3 \theta \right) \right]^{1/2} \right)
\end{align}
where
 the overall factor of 4 is due to the spin and valley degeneracy, we introduced $b = \frac{\sqrt{3}a}{2}$, and we remind that $\beta = D \gamma_0/(\gamma_1 \gamma_3)$ and $z = (\sqrt{3}/2) (\gamma^2_0/(\gamma_1 \gamma_3)) a k$.
It is instructive to first consider the limit when there is no displacement field, i.e. when $\beta = 0$. In this case,
\begin{align}
N_F^{\beta=0} &= 4\left(\frac{\gamma_1 \gamma_3}{\gamma_0^2 b}\right)^2 \int
\frac{z dz d \theta}{(2\pi)^2} \delta\left(\mu - \frac{ \gamma_1 \gamma_3^2}{\gamma_0^2}z \sqrt{1- 2 z \cos 3 \theta} \right)\\
&=4\frac{\gamma_1}{\gamma_0^2 b^2} \int
\frac{z dz d \theta}{(2\pi)^2} \delta\left(\frac{\gamma_0^2 \mu}{\gamma_1 \gamma_3^2} - z \sqrt{1 - 2 z \cos 3 \theta} \right).
\end{align}
Choosing $\mu$ such that the relevant $z$ are small, we can expand the square root to leading order in $z$.  Integrating over $\theta$, we obtain
\begin{align}
N_F^{\beta=0} &= \frac{4 \gamma_1}{\gamma_0^2 b^2} \int \frac{d^2z}{(2\pi)^2} \delta\left(\tilde \mu - z \left(1 - z \cos 3 \theta \right) \right) \\
&= \frac{8\gamma_1}{\gamma_0^2 b^2} \int \limits_{z_0}^{\infty} \frac{dz}{\left(2\pi \right)^2} \frac{z}{\sqrt{z^4-\left( \tilde \mu -z \right)^2}},
\end{align}
where $\tilde \mu = \gamma_0^2 \mu/\gamma_1 \gamma_3^2$ and $z_0 = \frac{1}{2} \left(\sqrt{1+4\tilde \mu} - 1\right)$.
The bounds on the integral must be handled with care, as the delta function has solutions only when $|\tilde \mu/z^2 -1/z|\leq 1$. Here, we assume that $\tilde \mu>1/4$, so this condition is satisfied for all $z>z_0$. In addition, $z = z_0$ is the only zero that occurs on the real axis. The remaining integral over $z$ is infrared convergent, as one can easily verify. It is formally ultraviolet divergent, but this is resolved by the presence of a quartic term in the dispersion that is not written in Eq. (\ref{dispersion}), but can be obtained by expanding Eq. (\ref{analytic_solution}) to fourth order in $k$.
Consequently, we conclude that
\beq
N^{\beta =0}_F \sim \frac{\gamma_1}{\gamma_0^2 b^2}
\label{nn_1}
\eeq

When the displacement field is finite,
\begin{align}
N_F &= 8\frac{\gamma_1 \gamma_3}{\beta \gamma_0^3 b^2} \int \frac{z dz d \theta}{(2\pi)^2} \delta \left( \frac{2 \gamma_0 \mu}{\beta \gamma_1 \gamma_3} - \left[1 + \frac{4\gamma_3^2}{\gamma_0^2 \beta^2}z^2\left(1 - \beta^2- 2 z \cos 3 \theta \right)\right]^{1/2} \right)
\end{align}
Making a substitution $z = \frac{\beta}{\sqrt{1-\beta^2}} \frac{\gamma_0}{\gamma_3} y$, we then obtain
\begin{align}
N_F &= 2\frac{\gamma_1}{\gamma_0 \gamma_3 b^2} \frac{\beta}{|1- \beta^2|} \int \frac{y dy d \theta}{(2\pi)^2} \delta \left( \frac{2 \gamma_0 \mu}{\beta \gamma_1 \gamma_3} - \left[1+ y^2\left(a - \frac{\gamma_0}{\gamma_3} \frac{\beta}{\left|1-\beta^2\right|^{3/2}} y \cos 3 \theta \right) \right]^{1/2} \right),
\end{align}
where the $a=1$  when $\beta<1$ and $a=-1$ when $\beta>1$.
We evaluated the integral both numerically and analytically (by neglecting the $y^3$ term under the $\delta$-function and verifying a posteriori that this is justified) and found that it is of order one.  Therefore, at a finite $\beta$,
\beq
N_F \sim \frac{\gamma_1}{\gamma_0 \gamma_3 b^2} \frac{\beta}{|1-\beta^2|}
\label{nn_2}
\eeq
Comparing with (\ref{nn_1}), we find that at $\beta > \gamma_3/\gamma_0$ (i.e., at $D > \gamma^2_3 \gamma_1/\gamma^2_0$),
the density of states is enhances compared to the case of zero displacement field.  For a generic $\beta = O(1)$, the enhancement is parametrically large:
\beq
N_F \sim N^{\beta =0}_F \frac{\gamma_0}{\gamma_3}
\label{nn_3}
\eeq
The ratio $\gamma_0/\gamma_3$ is around 10, so the density of states gets enhanced by order of magnitude.

\begin{figure}[h]
	\begin{center}
		\includegraphics[scale=.6]{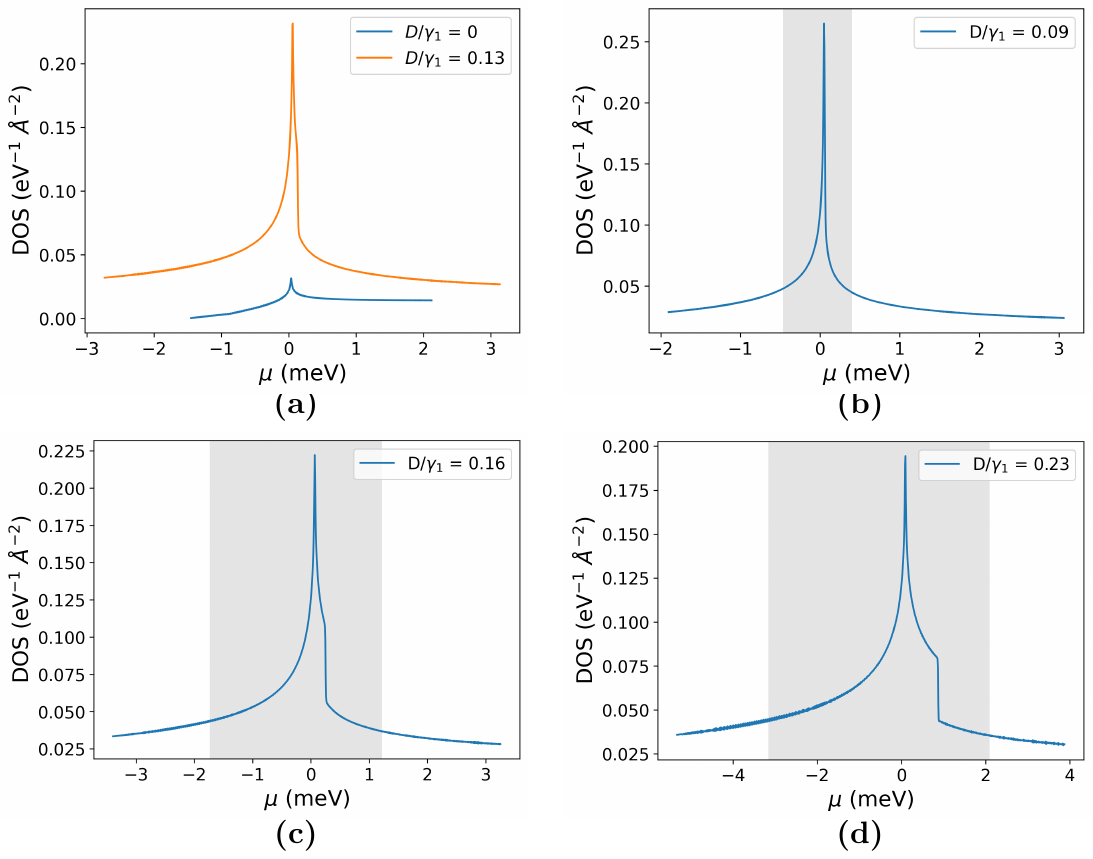}
		\caption{Numerically calculated density of states for the system for several different displacement fields. In (a), we demonstrate that the density of states is greatly enhanced in the presence of the displacement field. In (b)-(d), the shaded region represents where ordered states occur when the interaction is $SU(4)$ symmetric. In all figures, we have shifted the chemical potential so that the Van Hove singularity occurs at $\mu=0$.}
		\label{dos}
	\end{center}
\end{figure}

We show the results of numerical calculation of the density of states in Fig. \ref{dos}.
We emphasize that the enhancement is not restricted to the immediate vicinity of the Van Hove point, but instead holds in a wide range of fermionic densities.  The shaded regions in Fig. \ref{dos}b-d mark where ordered states appear (see below).   We see that in the presence of a displacement field, the density of states gets enhanced  everywhere in these ranges.  We believe this is the key reason why ordered states have been observed in the presence of a displacement field.

\subsection{Detection of the ordered states}
In this section, we outline how we detect spin and valley orders.
We remind that we use two interactions --  the density-density interaction between particles in the same valley, $U_1$, and the interaction between fermionic densities between different valleys, $U_2$. We write the interaction part of the Hamiltonian as
\begin{align}
H_{I} = \frac{1}{2}\sum_{\vb k,\vb p,\vb q, \tau, \tau',\alpha,\beta}\left( U_1\delta_{\tau \tau'} + U_2 \tau^x_{\tau \tau'} \right)c^{\dagger}_{\vb k + \vb q,\tau,\alpha}c^{\dagger}_{\vb p - \vb q,\tau',\beta}c_{\vb p,\tau',\beta}c_{\vb k,\tau,\alpha},
\label{nn_4}
\end{align}
where $\tau^{x}$ is the first Pauli matrix in valley space.

For complete analysis, the interaction terms in Eq. (\ref{nn_4}) has to be multiplies by coherence factors coming from the diagonalization of the $4\times 4$ quadratic form in Eq. (\ref{single_particle}). These phase factors generally have amplitudes and phases.  For particle-hole orders with $q=0$, which we will study, the phase factors for $c^\dagger_k$ and $c_k$ cancel out, and the amplitudes can be evaluated at $K$ and $K'$ and multiply $U_1$ and $U_2$ by some constants.  We assume that this renormalization is already absorbed into $U_1$ and $U_2$.

Previous studies have found that when $U_1 = U_2 = U$ and trigonal warping is absent, the Hamiltonian possesses $SU(4)$ symmetry~\cite{chichinadze2022,chichinadze2022letters}.
In this case, valley polarization and valley ferromagnetism develop simultaneously.
For non-equal $U_1$ and $U_2$, $SU(4)$ symmetry is broken and valley and spin orders develop separately.
We note in this regard that even if $U_1=U_2=U$ in Eq. (\ref{nn_4}), the irreducible $U_1$ and $U_2$ in the particle-hole channel are not the same due to different renormalizations in the particle-particle channel
~\cite{calvera2024nematicity,raines2024stoner}.
Here, we simply break the $SU(4)$ symmetry by hand, by treating $U_1$ and $U_2$ as two independent parameters.
For circular pockets (no trigonal warping), spin polarization develops first when $U_1>U_2$, and valley polarization develops first when $U_2>U_1$ (Ref. \cite{dong2023}).
In the presence of trigonal warping, it was argued~\cite{calvera2024nematicity} that spin order develops first even when $U_1 = U_2$.

In what follows we detect the instabilities towards spin and valley order within self-consistent ladder approximation, i.e., by summing up ladder series of diagrams in the spin or valley density-wave channel with the Green's functions dressed by the self-energy from the corresponding order.

We define the valley polarization order parameter as
\begin{align}
\Delta_{VP} = \frac{1}{N} \sum_{\vb k,\tau, \tau',\alpha,\beta} \langle c^\dagger_{k,\tau,\alpha} \tau^{z}_{\tau \tau'} \delta_{\alpha \beta} c_{k,\tau',\beta} \rangle
\end{align}
and the spin polarization order parameters as
\begin{align}
\Delta_{SP\tau} = \frac{1}{N} \sum_{\vb k,\alpha,\beta} \langle c^\dagger_{k,\tau,\alpha} \sigma^{z}_{\alpha \beta} c_{k,\tau,\beta} \rangle ,
\end{align}
where $N$ is the total number of particles and $\tau =1,2$.
The spin polarization order parameters can be combined into inter-valley ferromagnetism and inter-valley antiferromagnetism as
\begin{align}
\Delta_{FM} &= \Delta_{SP1} + \Delta_{SP2} \\
\nonumber \Delta_{AFM} &= \Delta_{SP1} -\Delta_{SP2}.
\end{align}
The two remain degenerate in our theory, but the degeneracy can be broken either by restoring $U_3$ or by incorporating the effects of spin-orbit coupling.

\section{Results}
\label{results}

In this section, we present our results for spin/valley orders.
In our numerical solutions to the order parameters, we vary both number density and displacement field.
For each density and field, we solve the self-consistent equations for the order parameters while also conserving particle number.
If multiple solutions are found, the solution that minimizes the free energy is taken.
For definiteness, we consider the case of hole doping, for which most experimental results have been obtained.

For the theoretical phase diagram we set $U_1 = U_2 = 242 \text{ eV\AA}^2$
to match the experimental data for the density at which there is a  transition from the normal state at the lowest densities to the quarter metal state at higher densities.

\subsection{Isotropic Dispersion}

\begin{figure}[h]
	\begin{center}
		\includegraphics[scale=.8]{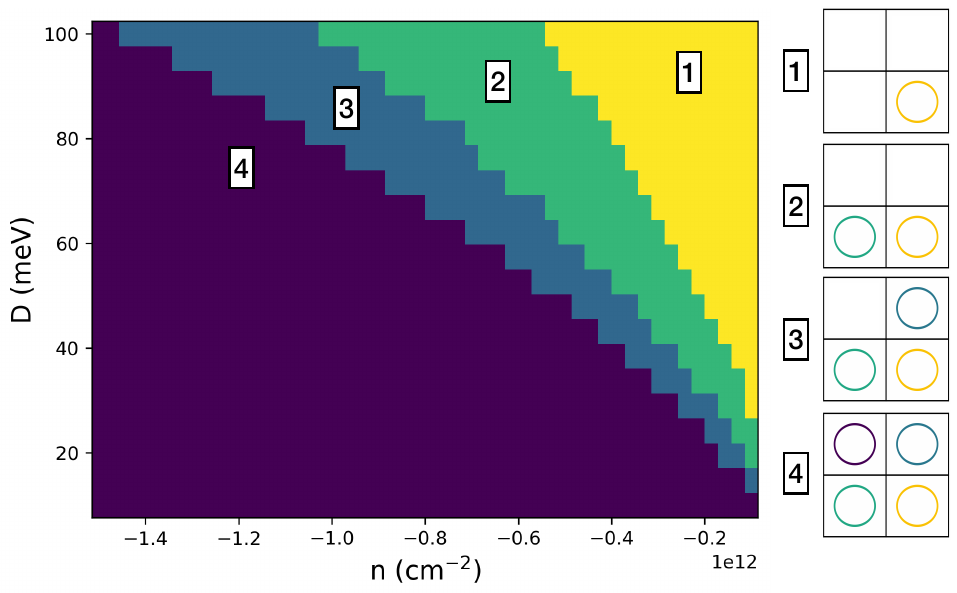}
		\caption{The phase diagram
of the system when no trigonal warping is present and the interaction is $SU(4)$ symmetric.
The ordered states are quarter-metal, half-metal, and three-quarter metal with full spin or valley and  spin and valley
polarizations. There are no partially polarized states. On the right we schematically plot the Fermi surfaces for each isospin, leaving
the individual isospins unlabeled due to $SU(4)$ symmetry. }
		\label{no_trigonal_warping}
	\end{center}
\end{figure}

We consider first the case where the interaction is $SU(4)$ symmetric and there is no trigonal warping.
To do this, we set $\gamma_3 = \gamma_4 = \delta = 0$. In agreement with earlier works~\cite{raines2024isospin,*raines2024unconventional}, we found a series of phase transitions with increasing doping, leading to quarter metal, half-metal, three-quarters metal and finally to a full metal. We present the phase diagram in Fig. \ref{no_trigonal_warping}.  There are no partially polarized states at any doping and displacement field.

\begin{figure}[h]
	\begin{center}
		\includegraphics[scale=.7]{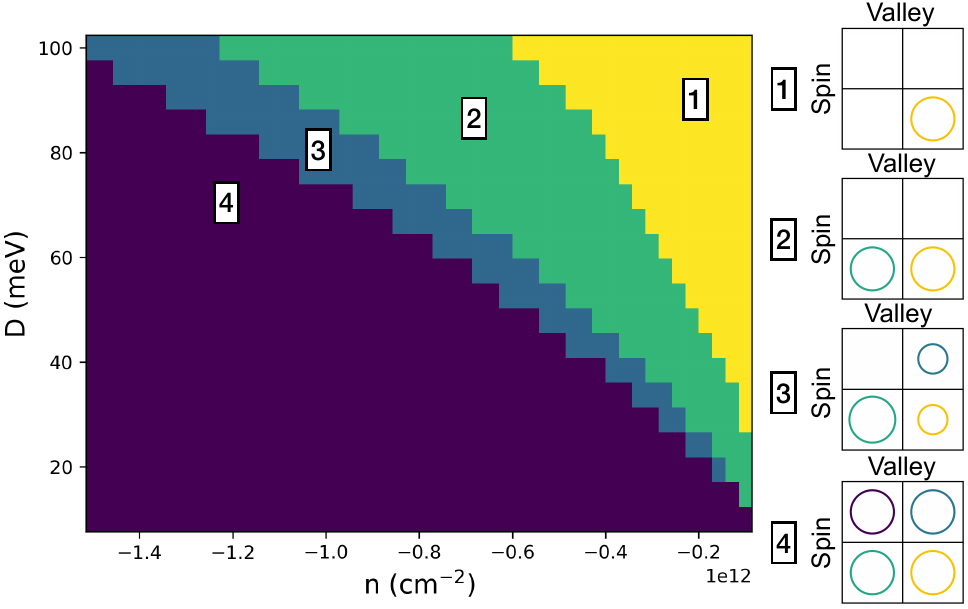}
		\caption{The phase diagram when $U_1 = \frac{5}{4} U_2$ and there is no trigonal warping. Relative to Fig. \ref{no_trigonal_warping}, the half metal phase takes up a larger region of the phase diagram, while the three-quarters metal phase has shrunk. In the panels to the right we now label the spin/valey components of the isospins, as they are no longer degenerate.}
		\label{asym_sp_no_trig}
	\end{center}
\end{figure}

\begin{figure}[h]
	\begin{center}
		\includegraphics[scale=.7]{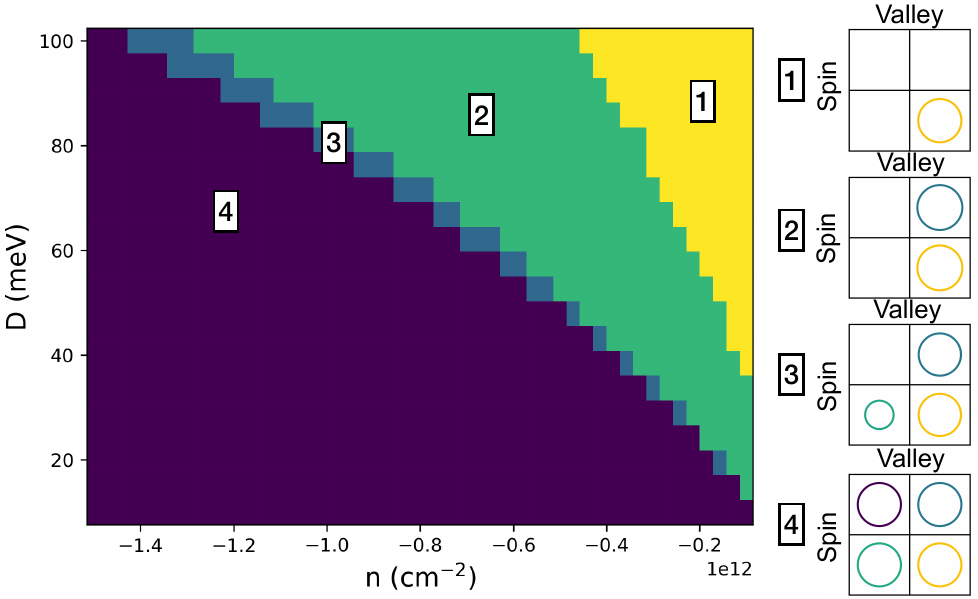}
		\caption{The phase diagram when $U_1 = \frac{3}{4} U_2$ in the absence of trigonal warping. The notations are the same as in Fig. \ref{asym_sp_no_trig}.}
		\label{asym_vp_no_trig}
	\end{center}
\end{figure}

When the $SU(4)$ symmetry is broken, either spin polarized or the valley polarized states become favored, depending on whether $U_1$ is larger than $U_2$ or $U_2 >U_1$.  For practical calculations we fixed $U_2 = 242 \text{ eV}\text{\AA}^2$ and
varied $U_1$.
 We show the results in Figs. \ref{asym_sp_no_trig} and \ref{asym_vp_no_trig}
 for  $U_1 = \frac{5}{4} U_2$ and $U_1 = \frac{3}{4} U_2$. In both the cases, we find results analogous to those in Refs. \cite{raines2024isospin,*raines2024unconventional}.
Namely,  the half metal phase expands and the three-quarters metal phase shrinks, especially when setting $U_1 =\frac{3}{4} U_2$. We  expect that in the limit of sufficiently strong asymmetry, the three-quarters metal phase would vanish.
We also note the presence of unequally sized Fermi surfaces, e.g., in the three-quarters metal phase, there are smaller and larger Fermi surfaces in three of the isospins.  However, the isospin that was unoccupied in the $SU(4)$ symmetric case remains unoccupied.

\subsection{Trigonally Warped Dispersion}

\begin{figure}[t]
	\begin{center}
		\includegraphics[scale=.7]{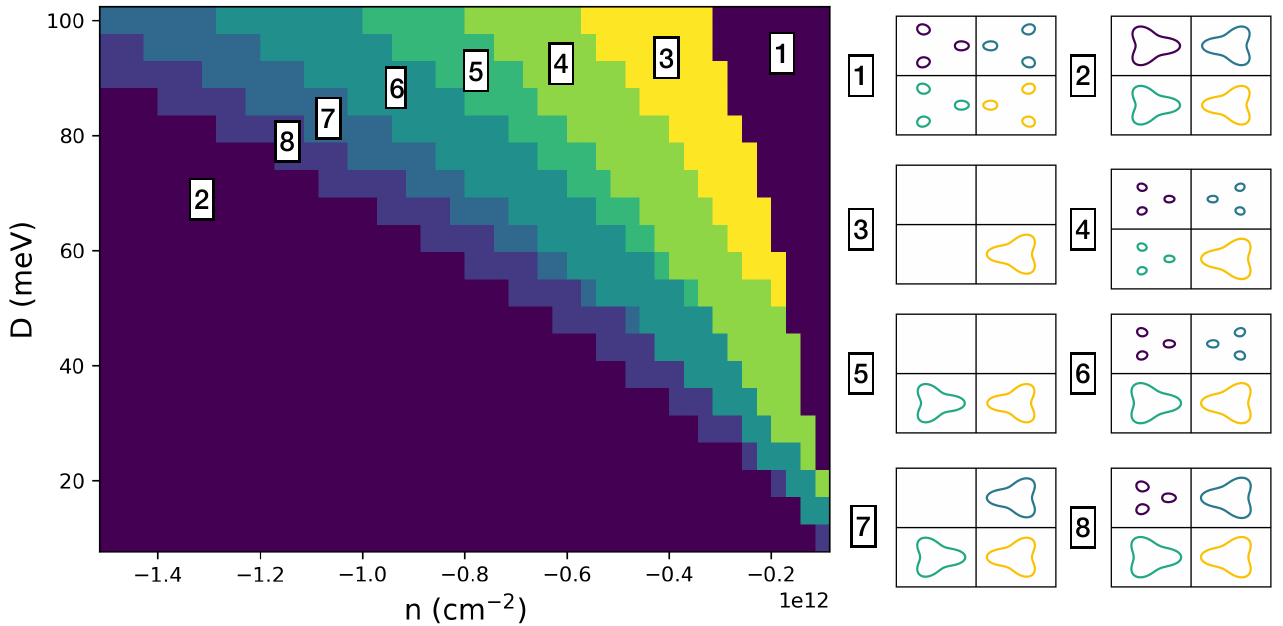}
		\caption{Phase diagram depicting the fully polarized and partially polarized states as both displacement field and particle number are varied. On the right are schematic depictions of the Fermi surface in each phase. As in the case of Fig. \ref{no_trigonal_warping}, we do not label the individual isospins since the interaction is $SU(4)$ symmetric.}
		\label{su4_trig_warping}
	\end{center}
\end{figure}

We next present our results for the case where the trigonal warping terms are nonzero. The phase diagram for $U_1 = U_2 = U$ is presented in Fig. \ref{su4_trig_warping}. For this choice of interactions, the order still does not distinguish between valley and spin polarization, yet the phase diagram is different from the one in Fig. \ref{no_trigonal_warping} for circular Fermi surfaces. The first noticeable difference is that at the lowest densities, there is now a full metal state. This is due to the fact that, in the presence of trigonal warping, the Van Hove singularity is located at a finite density of holes.
At the lowest hole doping, the density of states is not large enough to promote the formation of an ordered state for $U$ that we have chosen. Accordingly, the transition to the quarter metal state occurs at some finite density.

The other difference is the presence of partially polarized states in Fig. \ref{su4_trig_warping} (states labeled as 4, 6 and 8 in the right panel in the Figure). The general progression of states is the same as for circular Fermi surfaces, with the system evolving from the symmetric state to the quarter metal, half metal, three-quarters metal, and then full metal state. However, now in between the fully polarized states there are partially polarized states, in which there are now minority carriers in addition to the majority carriers. We note that each transition is still first order. That is, there is still a discontinuity in the isospin occupation at the transition between the partially polarized quarter metal and fully polarized half metal state and so on.

One can compare the phase diagram to that found in experiments (see Fig. \ref{pip}). The quarter metal state, labeled 3 in Fig. \ref{su4_trig_warping}, matches the IF$_{1}$ state. The state labeled 6 in Fig. \ref{su4_trig_warping} matches the PIP$_2$ state with two majority and two majority carriers. There is, however, no state matching PIP$_1$.

\begin{figure}[t]
	\begin{center}
		\includegraphics[scale=.6]{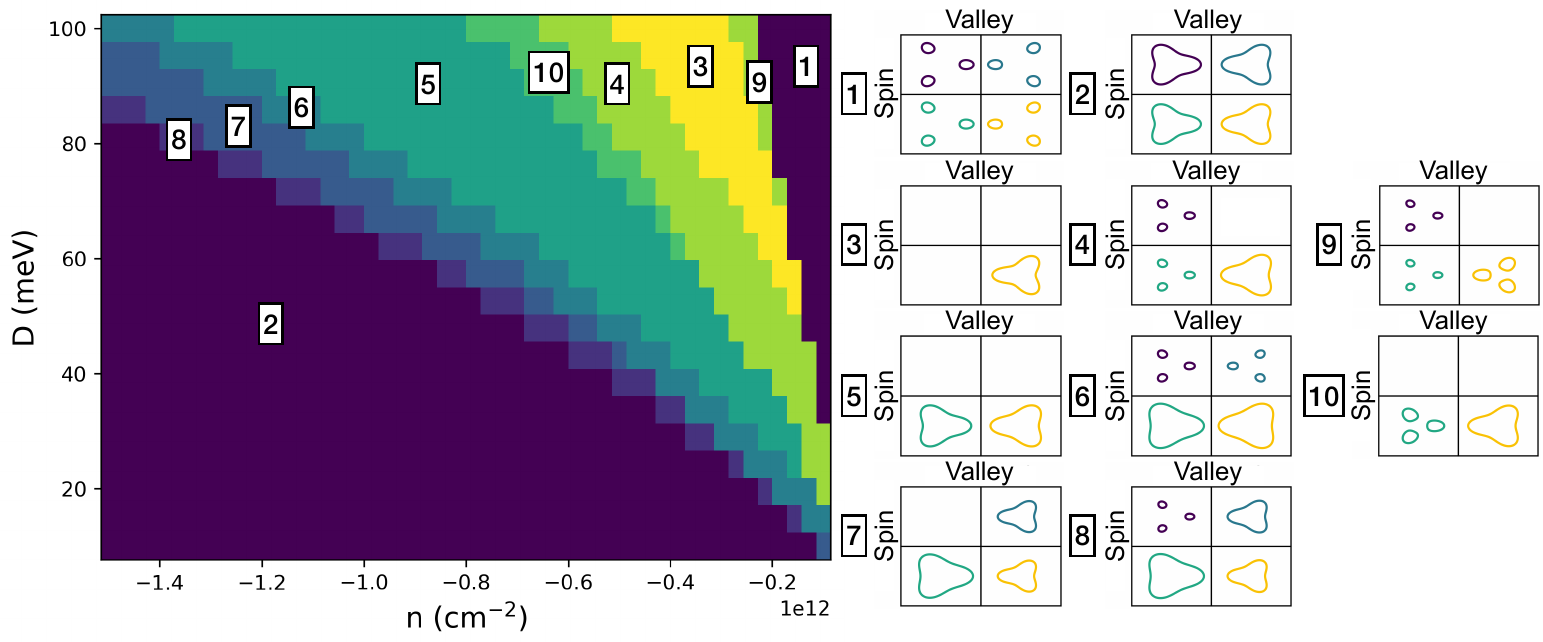}
		\caption{Phases when $U_{1}=\frac{5}{4} U_2$, where both number density and displacement field are varied. Here, we label each of the isospins, as the spin and valley polarized states are no longer degenerate with each other. }
		\label{sp_phase}
	\end{center}
\end{figure}

We next present the results for the anisotropic interaction.  As before, we fix the value of $U_2$ and vary $U_1$.
We  present the case  $U_1=\frac{5}{4}U_2$ in Fig. \ref{sp_phase}.  For this case, a half-metal state is spin polarized, and a quarter-metal state  has an additional valley polarization for the spin components forming the Fermi surfaces.
We see  that a number of new states arise relative to the  case $U_1 = U_2$. These are states 4, 9 and 10 on the right panel in Fig. \ref{sp_phase}. In each of these new states, there is an unequal occupation of the minority isospins.
For instance, in the state labeled as 10 the system is fully spin polarized but only partially valley polarized. This matches the PIP$_1$ state found in experiments.  The state labeled as 6 matches the PIP$_2$ state.

We also see that generally the width of the half metal state grows substantially compared to the case of equal interactions. This is behavior is expected, as is the same as for the isotropic dispersion and circular Fermi surfaces. The transitions between different states are again all first order.
\begin{figure}[t]
	\begin{center}
		\includegraphics[scale=.6]{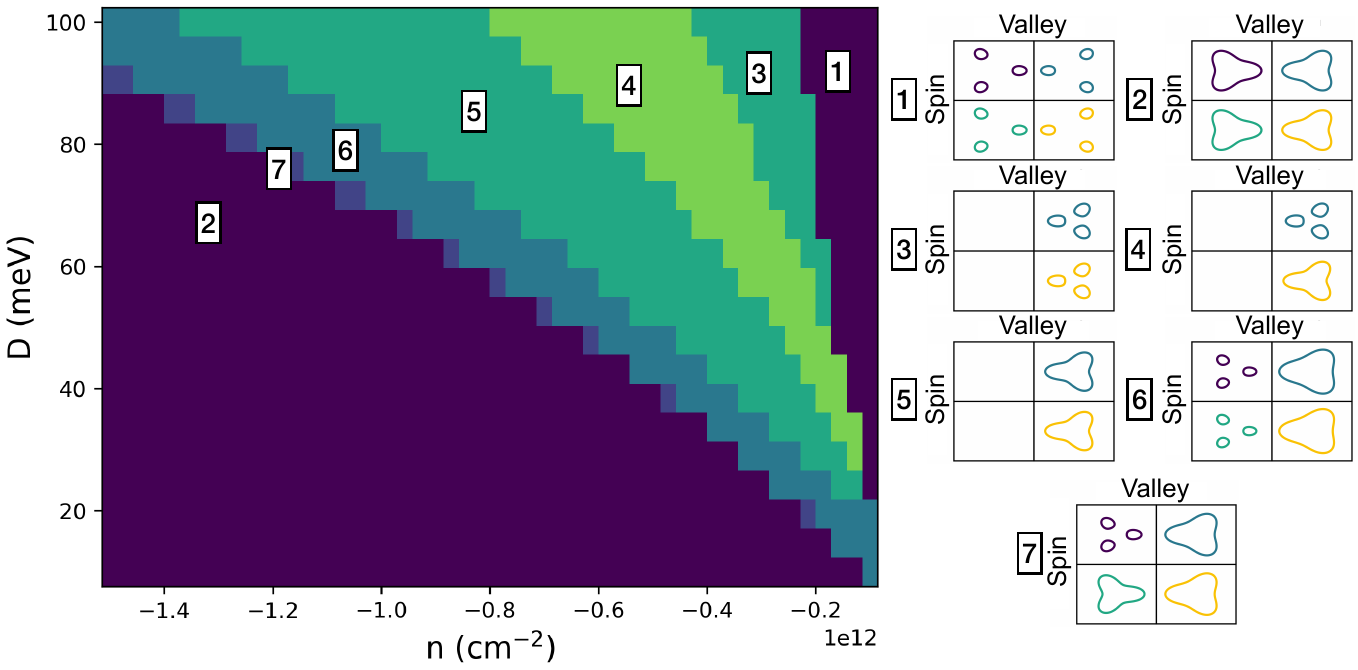}
		\caption{Phase diagram when $U_1 = \frac{3}{4}U_2$, where both number density and displacement field are varied. }
		\label{vp_phase}
	\end{center}
\end{figure}

In Fig. \ref{vp_phase} we present the results
for $U_1 = \frac{3}{4} U_2$. Now the half-metal state is the state with full valley polarization and no spin order and the quarter-metal state has additional full spin polarization within the valley crossing the Fermi level. Besides these fully polarized states, we again find a set of partly polarized states, some matching the ones observed in the experiments. Specifically, the state labeled as 10 matches the PIP$_1$ state, and the state labeled as 6 again matches the PIP$_2$ state.

\subsubsection{Number of pockets for minority carriers}

There is one aspect in which our results differ from the data \cite{zhou2022bbg,zhang2023,holleis2025}
Namely, the PIP$_1$ and PIP$_2$ states contain one small and one large Fermi pocket and two small and two large Fermi pockets. Our partially polarized states contain small and large Fermi pockets, but the small pocket actually consists of three small Fermi surfaces. In our theory, this comes about because partial spin or valley order effectively moves the Fermi level of minority carriers towards that at smaller dopings, where the Fermi surface consists of three small pockets
 (see Fig. \ref{fermi_surfaces}). In principle, it is possible that  both  majority and minority carriers form single Fermi pockets, but we didn't obtain this within our model.   Another possibility is a nematic order, which breaks $C_3$ symmetry and reconstructs three small pockets into either a single one or into two pockets.  To search for a nematic order, one has to keep the coherence factors in the interactions and search for a particle-hole order with an angular dependent prefactor.  This will be the subject for subsequent study.

\begin{figure}[t]
	\begin{center}
		\includegraphics[scale=.6]{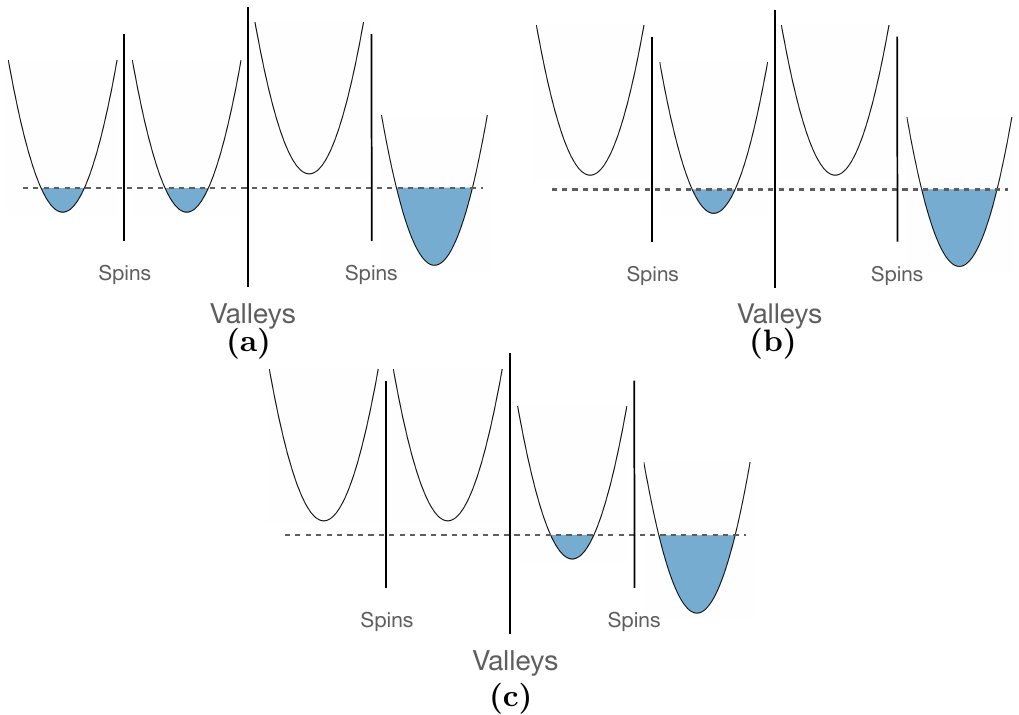}
		\caption{Schematic depiction of the phases unique to the interaction asymmetric case when $U_1>U_2$ (a-b) and when $U_2>U_1$ (c). From Fig. \ref{sp_phase} we depict phases 4 and 9 in (a) and phase 10 in (b). In (c) we depict phase 4 from Fig. \ref{vp_phase}. }
		\label{isospin_occupation}
	\end{center}
\end{figure}

\section{Conclusions}
\label{conclusions}
In this work, we have systematically analyzed the effects of trigonal warping and asymmetry of the interactions on the structure of $q=0$ ordered states in BBG. We used a realistic dispersion for the BBG bands and analyzed  the ordered spin and valley fully and partly polarized states, induced by the interaction terms $U_1$ and $U_2$, involving fermionic densities within a given valley and in different valleys. By computing the particle-hole spin and valley (pseudospin) ordered states within the ladder approximation and comparing the energies of different ordered states, we determined the phase diagram for the system for several different sets of parameters.

We demonstrated that the partially polarized states, similar to those observed in the experiments, arise strictly from the presence of the trigonal warping term in the dispersion. In its absence, the states are always fully polarized and there are no minority carriers present in any ordered state. Using the value of the trigonal warping term found in \textit{ab initio} calculations, we found that partially polarized states make up a substantial portion of the phase diagram.

In addition to the trigonal warping, we also considered the effects of non-equal interactions $U_1$ and $U_2$. We considered both $U_1 > U_2$ and $U_1<U_2$ and determined the phase diagram in both instances. Predictably, as one interaction gets larger than the other, the spin or valley polarized phase grow relative to the symmetric case.  The structures of  the ordered states at  $U_1 > U_2$ and $U_1<U_2$  are not identical, but in both cases we found two partially ordered states closely matching experimentally observed ``almost" half-metal and ``almost" quarter-metal states  with both large and small Fermi surfaces.

\section{Acknowledgments}

We acknowledge with thanks useful discussions with
 E. Berg, Z. Dong, S. Nadj-Perge, Z. Raines, A. Seiler, T. Weitz  and A. Young. This work was supported by U.S. Department
of Energy, Office of Science, Basic Energy Sciences,
under Award No. DE-SC0014402.

\bibliography{bbg_bib}

\end{document}